\documentclass[11pt, a4paper]{article}
\pdfoutput=1
\usepackage{jheppub}
\usepackage{graphicx}
\usepackage{epsfig,amsmath,amsthm}
\usepackage{epstopdf}
\usepackage{youngtab}

\newcommand{\be}{\begin{equation}}
\newcommand{\ee}{\end{equation}}
\newcommand{\bea}{\begin{eqnarray}}
\newcommand{\eea}{\end{eqnarray}}
\def\bse{\begin{subequations}}
\def\ese{\end{subequations}}
\newcommand{\IR}{\mathbb{R}} \newcommand{\IC}{\mathbb{C}}
 
\def\IZ{\relax\ifmmode\hbox{Z\kern-.4em Z}\else{Z\kern-.4em Z}\fi}

\def\IP{\mathbb{P}}

\newcommand{\non}{\nonumber \\}

  \def\eps{\epsilon}

 \def\sig{\sigma}

\def\bi{\begin{itemize}} \def\ei{\end{itemize}}

\def\({\left(} \def\){\right)}
\def\[{\left[} \def\]{\right]}
\def\<{\left<} \def\>{\right>}

\def\om{\omega}

\title{Perturbative gauge theory and $2+2=4$}

\author{Barak Kol and Ruth Shir \\
{\it Racah Institute of Physics, Hebrew University, Jerusalem 91904, Israel} \\
{\tt barak.kol,ruth.shir@mail.huji.ac.il}
}

\abstract{The group $S_4$ of permutations on four elements has an irreducible representation corresponding to the partition $4=2+2$. This representation appears in several different mathematical contexts: the Jacobi identity of Lie algebras; the Schouten identity for spinors; differences and the cross ratio in the projective plane; the Grassmannian space of 2-planes in linear algebra; the Riemann tensor in differential geometry; and more.   We observe that all these occurrences are connected through perturbative non-abelian gauge theory, thereby acting as a leitmotif.} 

\begin{document}
\maketitle

\section{Introduction}

Recently we studied the space of color structures for tree-level scattering of $n$ gluons in pure Yang-Mills \cite{color}, known to be $(n-2)!$ dimensional, and we presented its transformation properties under $S_n$, the group of permutations of external legs, by recognizing a linkage with the notion of the cyclic Lie operad.

During that investigation and the study of the field of scattering amplitudes, see the reviews \cite{ManganoParke1990,BernDixonKosower1996,DixonRev2011,ElvangHuang2013,DixonRev2013} and references therein, we observed a recurring theme which is the subject of the current paper. 

\section{The $2+2$ irrep of $S_4$}

$S_4$, the group of permutations on 4 objects, has an irreducible representation (irrep) associated with the partition of 4 into $2+2=4$, or equivalently the Young diagram \be
 \yng(2,2)
 \label{Young}
\ee
In this section we construct this irrep and discuss it.

The representation can be constructed as follows. Consider a function $t=t_\sig$, where $\sig \in S_4$ sends $(1 2 3 4) \to (i j k l)$, satisfying the following symmetries \bea
 t_{j i k l} &=& - t_{i j k l} \non
 t_{k l i j} &=&   t_{i j k l}
 \label{construct} \\
 0 &=& t_{i j k l} + t_{k i j l} + t_{j k i l} 
 \label{construct2}
 \eea

The first two identities imply that $t$ is anti-symmetric with respect to interchanging either the first pair of indices or the second, and  it is symmetric with respect to swapping the pairs. These eight symmetries reduce the dimension of possible $t_\sig$ from 24 to 3. The third identity implies that the totally anti-symmetric part vanishes. Hence the dimension of this representation is exactly 2, and it turns out to be irreducible.\footnote{
  This can be shown for example by noting that the representation is free of both a totally symmetric and a totally anti-symmetric part, and inspection of the irreps of $S_4$ shows that such a 2d representation is irreducible and unique.} 

Equivalently this representation can be constructed by identities similar to (\ref{construct},\ref{construct2}) only without the minus sign in the first identity. In this case the third identity serves to project out the totally symmetric component. Following the same reasoning as before the dimension is 2, it is the same irrep and the two representations are in fact equivalent.

By construction this representation is neither fully symmetric nor fully anti-symmetric and as such it constitutes an intermediate type of symmetry.

\section{Manifestations}

In this section we shall see several appearances of the $2+2$ symmetry type, which are apparently independent, yet are interconnected through scattering amplitudes. 

\subsection{Lie algebra and Jacobi identity}

The Jacobi identity is the defining property of a Lie algebra. Its standard form is $\[ \[ T^a,\, T^b\], T^c\] + \[ \[ T^c,\, T^a\], T^b\] +\[ \[ T^b,\, T^c\], T^a\] =0$, where $T^a, T^b, T^c$ are any 3 generators.  In terms of the algebra structure constants $f^{a b c}$ \footnote{
 They are defined by $\left[ T^a,\, T^b \right] = i\, f^{a b c}\, T^c$, where $T^a$ are the (Hermitian) generators (see for example the textbook \cite{PeskinSchroeder} p. 490 eq. (15.44)). Their normalization is immaterial for current purposes.}
 it is given by
  \be
f^{a b x}\, f^{x c d} + f^{c a x}\, f^{x b d}+ f^{b c x}\, f^{x a d}=0 ~.
\label{Jacobi}
\ee
This form stresses the action of $S_4$ on the identity (as usual one uses the Killing form to raise indices in $f^{a b c}$), while the former makes only an $S_3$ manifest. In fact, we may define \be
t^J_{i j k l} := f^{i j x}\, f^{x k l}
\label{def:tJ}
\ee
$t^J$ has the same symmetries as in (\ref{construct}) where the lack on a totally anti-symmetric component (\ref{construct2}) is guaranteed exactly by the Jacobi identity.\footnote{
For any choice of distinct $i, j, k, l$ we get the $2+2$ representation. Alternatively the whole tensor $t^J$ transforms according to the $2+2$ Young diagram (\ref{Young}).} 
 Hence $t^J$ has $2+2$ symmetry.

We would like to mention a diagrammatical form of the Jacobi identity, namely \be
\parbox{25mm}{\includegraphics[scale=1]{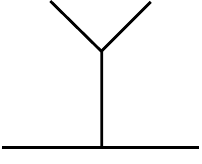}} = \parbox{25mm}{\includegraphics[scale=1]{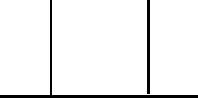}} - \parbox{25mm}{\includegraphics[scale=1]{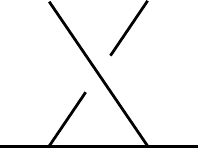}} 
\label{Jacobi-diag}
\ee
where each vertex is oriented clockwise and it represents one factor of $f^{a b c}$. This form was presented first in \cite{Cvitanovic-etal1981}.

\subsection{Spinors and Schouten identity}

Given 4 spinors, denoted $1, 2, 3, 4$ we can define \be
t^S_{1 2 3 4} := \left<1\, 2 \right>\, \left< 3\, 4 \right>
\label{def:tS}
\ee
where $\left< \lambda \, \mu \right> := \eps_{\alpha \beta}\, \lambda^\alpha \, \mu^\beta$.
 Again $t^S$ has the symmetries (\ref{construct}). This time the absence of a totally anti-symmetric part (\ref{construct2}) becomes the Schouten identity \be
 0 =  \left<1\, 2 \right>\, \left< 3\, 4 \right> +  \left<2\, 3 \right>\, \left< 1\, 4 \right> +  \left<3\, 1 \right>\, \left< 2\, 4 \right> ~.
 \label{Schouten}
 \ee
 It has the geometrical interpretation that the 4-volume (namely the totally anti-symmetric part) of 4 2d spinors, must vanish. It is seen that $t^S$ has $2+2$ symmetry. 

This identity is of course central to the manipulation of helicity spinors for scattering amplitudes, and in particular for demonstrating that the permutation-dependent part of the Parke-Taylor formula \cite{ParkeTaylor86}, namely \be
\[ \left<1\, 2 \right>\, \dots \left<n-1\, n \right>\, \left<n\, 1 \right> \]^{-1}
\ee
 satisfies the Kleiss-Kuijf relations \cite{KleissKuijf88} (which originate with the symmetries of color structures).

\subsection{Projective plane and Gross-Mende scattering equations}

Cachazo-He-Yuan (CHY) \cite{CachazoHeYuan13} presented a very interesting expression for arbitrary tree-level scattering amplitudes in both gauge theory and gravity, with no reference to helicity spinors, and none to the number of negative helicities or the space-time dimension. The CHY expression depends on sets of complex numbers, $\sig_i$, where the index $i$ runs over external legs, which are solutions of the Gross-Mende scattering equations.

If we now define \be
t^{\sig}_{i j k l} = (\sig_i-\sig_j) (\sig_k-\sig_l)
\label{def:tsig}
\ee
then again the symmetries (\ref{construct}) are manifest while the absence of the anti-symmetric part (\ref{construct2}) is now a matter of straightforward algebra. Hence $t^\sig$ has $2+2$ symmetry.

The permutation-dependent part of the CHY expressions is \be
\[ (\sig_1 -\sig_2) \dots (\sig_{n-1} - \sig_n) (\sig_n - \sig_1) \]^{-1}
\label{PTsector}
\ee
Just as for Parke-Taylor also here the 2+2 symmetry is sufficient to prove that this expression satisfies the Kleiss-Kuijf relations.

The projective plane $\IC \IP^1$ is directly related to the spinors of the previous subsection, being their projective version. Under this mapping the product $\left<1\, 2 \right>$ is mapped to $\sig_1-\sig_2$ which is used here.

{\bf Geometrical realization}. An interesting geometrical realization of the space of tree-level color structures $TCS_n$ was presented in \cite{Getzler94} through the $n-3$ homology of the space ${\cal M}_{0,n}$ of $n$ marked points on $\IC \mathbb{P}^1$ \be
TSC_n \simeq sgn_n \otimes H_{n-3}({\cal M}_{0,n})
\ee
 where $sgn_n$ is the sign representation of $S_n$, and $n-3$ is the top homology due  to the 3d projective symmetry, namely M\"obius transformations.

The cohomology of ${\cal M}_{0,n}$ can be constructed from Arnold's 1-forms \be
\om_{ij} = \frac{1}{2 \pi i} d\log \( \sig_i - \sig_j \)
\label{Arnold}
\ee
which satisfy the following relations \bea
 \om_{i j} &=& \om_{j i} \non
 0 &=& \om_{i k} \wedge \om_{j k} + \om_{j i} \om_{k i} + \om_{k j} \om_{i j}
 \eea
where here there is no summation on repeated indices. This means that $\om_{i k} \wedge \om_{j k}$ is in the $2+1$ representation of $S_3$ which is 2d and closely related to the $2+2$ representation of $S_4$. In fact, the $2+2$ representation has as its kernel the Klein subgroup $K \subset S_4$, which is isomorphic to $\IZ_2 \times \IZ_2$, and is hence also a representation of the quotient $S_4/K \simeq S_3$.

Altogether this suggests a connection between the CHY amplitudes, the $2+2$ representation and the space of color structures.

{\bf The cross ratio}. There is another relevant quantity associated with 4 points in $\sig_i \in \IC \IP^1, ~i=1,\dots,4$, namely the cross-ratio \be
 (\sig_1,\sig_2;\sig_3,\sig_4) = \frac{(\sig_1-\sig_3)(\sig_2-\sig_4)}{(\sig_2-\sig_3)(\sig_1-\sig_4)}
\label{xratio}
\ee
which is constructed of course to be invariant under projective transformations. Defining \be
t^x_{i j k l} = \log  \left| (\sig_i,\sig_j;\sig_k,\sig_l) \right|
\label{def:tx}
\ee
we find that $t^x$ satisfies the symmetries (\ref{construct}) while the identity (\ref{construct2}) is a result of the algebraic identity \be
  (\sig_1,\sig_2;\sig_3,\sig_4)\,  (\sig_3,\sig_1;\sig_2,\sig_4) \,  (\sig_2,\sig_3;\sig_1,\sig_4) = -1 ~.
\label{x-id}
\ee
Hence $t^x$ has $2+2$ symmetry.

The definition of the cross ratio generalizes to points in a higher dimensional space $\sig_i \in \IR^k, ~i=1,\dots,4$ through the replacement $\sig_i - \sig_j \to \left| \sig_i -\sig_j \right|$. In this way one defines the conformal cross ratio which is invariant under conformal transformations, see \cite{Basso:2013vsa} for a recent application. It, in turn, can be used to define a generalized $t^x$ satisfying also the $2+2$ symmetry.

\subsection{The common Grassmannian}  

The quantities $t^S,\, t^\sig$ which satisfy the original relations (\ref{construct},\ref{construct2}) share a common quadratic-like structure. In both cases we can write \be
t_{i j k l} = s_{i j}\, s_{k l}
\label{t-from-s}
\ee
where \be
s_{i j} = - s_{j i} ~.
\label{s-asym}
\ee
For the spinor algebra this is done through $s^S_{i j}:=\left< i\, j \right>$ while for the $\IC \IP^1$ algebra this is done through $s^\sig:=\sig_i -\sig_j$. These assignments are consistent with the projective mapping from spinors to $\IC \IP^1$ mentioned after (\ref{PTsector}). The Jacobi identity is also quadratic in a similar way, only there the squaring of $f^{i j x}$ (which is the candidate for $s_{i j}$) is accompanied also by summation over $x$. 

Note that the relations (\ref{s-asym},\ref{t-from-s},\ref{construct2}) can be interpreted geometrically to mean that $s_{i j}$ belongs to the Grassmannian $Gr(2,n)$, the space of 2-planes in $n$-dimensional space. Indeed the first identity implies that $s_{i j}$ is a bivector, while the second implies that $s \wedge s =0$ and hence $s = u \wedge v$ for some vectors $u,v$, that is $s$ represents a 2-plane.

In this sense the 2-Grassmannian is seen to underline all these previous cases. It may very well be related to the appearance of Grassmannians in scattering amplitudes in another context \cite{ArkaniHamedCachazoCheung2009,PositiveGrassmannian2012}.

\section{Related structures}

There are some other mathematical structures which have the 2+2 symmetry, and as such could possibly be related to scattering amplitudes.

\subsection{The Riemann curvature tensor}

The Riemann curvature tensor \be
R_{i j k l}
\label{Riemann}
\ee
has exactly the $2+2$ symmetries, namely (\ref{construct},\ref{construct2}). This symmetry is likely playing a role for gravitational scattering amplitudes.

\subsection{Solving the quartic and Galois theory}

The treatment of polynomial equations of degree $k$ is known to be closely related to $S_k$. Galois theory is exactly that mathematical theory which studies the symmetries of various expressions in the roots, and uses them to seek a solution.  This connection with permutation symmetries including the $2+2$ symmetry suggests that Galois theory might have a part to play in scattering amplitudes.

\subsection{Links and skein relations}

A relation similar to (\ref{construct2}) appears in the theory of knots and links. Considering a link with 4 loose ends there are 3 possible ways to reconnect them into pairs. A skein relation is a general term for a relation between the 3 corresponding invariants (its specific form depends of course on the chosen invariant).  In that sense it is similar to (\ref{construct2}) which also relates the 3 possible ways to divide the 4 indices into two pairs.

\section{Summary}

We observed that the $2+2$ symmetry-type occurs in several mathematical objects which are quite independent of each other, including Lie algebras through the Jacobi identity; spinors through the Schouten identity; the projective plane $\IC \IP^1$ in several forms; and the 2-Grassmannian.  Yet all these structures appear in the study of perturbative gauge theory (Yang-Mills), and are in fact interconnected there, thereby serving as a recurring theme, or leitmotif.

We noted that other mathematical structures share this symmetry including the Riemann curvature tensor, the Galois theory for solving the quartic and higher order equations, and skein relations for knots and links. The Riemann tensor is surely related to gravitational scattering amplitudes. The formal similarity suggests the possibility that Galois theory or skein relations would be found to take part in the theory as well.
 
\subsection*{Acknowledgments}

It is a pleasure to thank L. Mason, R. Monteiro and other members of the Oxford group for useful comments on a presentation of this work and for hospitality. BK is grateful to the organizers of the Kallosh / Shenker fest at Stanford and to N. Arkani-Hamed and J. Maldacena for hospitality at the Princeton Institute for Advanced Study during the initial stages of this work. 

This research was supported by the Israel Science Foundation grant no. 812/11 and it is part of the Einstein Research Project "Gravitation and High Energy Physics", which is funded by the Einstein Foundation Berlin.


\begin{thebibliography}{99}


\bibitem{color} 
  B.~Kol and R.~Shir,
  ``Color structures and permutations,''
  arXiv:1403.6837 [hep-th].

\bibitem{ManganoParke1990} 
  M.~L.~Mangano and S.~J.~Parke,
  ``Multiparton amplitudes in gauge theories,''
  Phys.\ Rept.\  {\bf 200}, 301 (1991)
  [hep-th/0509223].

\bibitem{BernDixonKosower1996} 
  Z.~Bern, L.~J.~Dixon and D.~A.~Kosower,
  ``Progress in one loop QCD computations,''
  Ann.\ Rev.\ Nucl.\ Part.\ Sci.\  {\bf 46}, 109 (1996)
  [hep-ph/9602280].

\bibitem{DixonRev2011} 
  L.~J.~Dixon,
 ``Scattering amplitudes: the most perfect microscopic structures in the universe,''
  J.\ Phys.\ A {\bf 44}, 454001 (2011)
  [arXiv:1105.0771 [hep-th]].
  
\bibitem{ElvangHuang2013} 
  H.~Elvang and Y.~-t.~Huang,
  ``Scattering Amplitudes,''
  arXiv:1308.1697 [hep-th].
  
\bibitem{DixonRev2013} 
  L.~J.~Dixon,
  ``A brief introduction to modern amplitude methods,''
  arXiv:1310.5353 [hep-ph].

\bibitem{PeskinSchroeder}
M.~E.~Peskin and D.~V.~Schroeder,
``An Introduction to Quantum Field Theory,''
{\it Westview Press} 1995.

\bibitem{Cvitanovic-etal1981} 
  P.~Cvitanovic, P.~G.~Lauwers and P.~N.~Scharbach,
  ``Gauge Invariance Structure of Quantum Chromodynamics,''
  Nucl.\ Phys.\ B {\bf 186}, 165 (1981).

\bibitem{ParkeTaylor86} 
  S.~J.~Parke and T.~R.~Taylor,
  ``An Amplitude for $n$ Gluon Scattering,''
  Phys.\ Rev.\ Lett.\  {\bf 56}, 2459 (1986).
 
 \bibitem{KleissKuijf88} 
  R.~Kleiss and H.~Kuijf,
  ``Multi - Gluon Cross-sections and Five Jet Production at Hadron Colliders,''
  Nucl.\ Phys.\ B {\bf 312}, 616 (1989). 
  
\bibitem{CachazoHeYuan13} 
  F.~Cachazo, S.~He and E.~Y.~Yuan,
  ``Scattering of Massless Particles in Arbitrary Dimension,''
  arXiv:1307.2199 [hep-th].
  
\bibitem{Getzler94}
E.~Getzler,
 ``Operads and moduli spaces of genus 0 Riemann surfaces,''
 The moduli space of curves (Texel Island, 1994), 199-230,
Progr.\ Math.\ 129, Birkh\"auser Boston, Boston, MA (1995)
[alg-geom/9411004]. 
  
  
\bibitem{Basso:2013vsa} 
  B.~Basso, A.~Sever and P.~Vieira,
  ``Spacetime and Flux Tube S-Matrices at Finite Coupling for N=4 Supersymmetric Yang-Mills Theory,''
  Phys.\ Rev.\ Lett.\  {\bf 111}, no. 9, 091602 (2013)
  [arXiv:1303.1396 [hep-th]].
  
\bibitem{ArkaniHamedCachazoCheung2009} 
  N.~Arkani-Hamed, F.~Cachazo and C.~Cheung,
  ``The Grassmannian Origin Of Dual Superconformal Invariance,''
  JHEP {\bf 1003}, 036 (2010)
  [arXiv:0909.0483 [hep-th]].

\bibitem{PositiveGrassmannian2012} 
  N.~Arkani-Hamed, J.~L.~Bourjaily, F.~Cachazo, A.~B.~Goncharov, A.~Postnikov and J.~Trnka,
  ``Scattering Amplitudes and the Positive Grassmannian,''
  arXiv:1212.5605 [hep-th].
  
\end{thebibliography}
\end{document}